\documentclass[aps,prl,twocolumn,amsmath,amssymb,nofootinbib,floatfix]{revtex4-1}

\usepackage{amsmath}
\usepackage{amssymb}
\usepackage{amsthm}
\usepackage[dvipdf]{color}
\usepackage{graphicx}
\usepackage{dcolumn}
\usepackage{bm,subfigure} 
\usepackage{xcolor,colortbl}
\usepackage[normalem]{ulem}

\DeclareMathOperator{\arcsinh}{arcsinh}
\DeclareMathOperator{\tr}{tr}

\newcommand{\KS}[1] {{\color{blue}#1}}

\begin{document}
\title{Fractional solitons in non-Euclidian elastic plates}

  \author{Kai Sun}
  \author{Xiaoming Mao}

 \affiliation{
 Department of Physics,
  University of Michigan, Ann Arbor, 
 MI 48109-1040, USA
 }

\begin{abstract} 
We show that minimal-surface non-Euclidean elastic plates share the same low-energy effective theory as Haldane's dimerized quantum spin chain.  As a result, such elastic plates  support fractional excitations, which take the form of charge-$1/2$ solitons between degenerate states of the plates, in strong analogy to their quantum counterpart. These fractional solitons exhibit properties similar to fractional excitations in quantum fractional topological states, including deconfinement and braiding, as well as unique new features such as holographic properties and diode-like nonlinear response, demonstrating great potentials for applications as mechanical metamaterials.
%We find that 2D minimal-surface plates shares the same low-energy effective theory as Haldane's dimerized quantum spin chain (a compact Sine-Gordon with an $Z_2$ symmetry), and thus low-energy excitations in these elastic systems are composed of integer and fractional solitons in analogy to $Z_2$ spin liquids and Haldane's dimerized spin chains.
\end{abstract}
\maketitle

\noindent{\it Introduction}---
The analogy between quantum and classical physics plays an important role in the history of many-body physics.  
%In the study of many-body physics, the interplay between quantum and classical systems play an important role. 
For example, in the early development of quantum topological states, concepts of classical topological defects (e.g., vortices and solitons) have been crucial to the theoretical understanding of fractional excitations in fractional quantum Hall systems~\cite{Laughlin1983} and quantum spin chains~\cite{Haldane1982,Haldane1983}. A more recent example is the duality between topological defects in elasticity and the newly proposed exotic quantum excitations, fractons, in tensor gauge theories~\cite{Pretko2018}.
%\sout{concepts developed from  classical topological excitations have been crucial in the discovery of quantum fractional topological states and fractional excitations}.  
Conversely,  quantum topological states of matter inspired the blossoming new field of topological mechanics~\cite{Kane2014,Prodan2009,Nash2015,Wang2015,Susstrunk2015,Paulose2015,Paulose2015a,Rocklin2016,Rocklin2017,Zhou2018,Zhou2019,Zhang2018,Lubensky2015,mao2018maxwell,Sun2020}. So far, mechanical analogs have only been achieved for integer quantum topological states, but not yet the more exotic fractional ones.

In a typical quantum system, excitations are usually composed of integer numbers of fundamental building blocks (quanta). However, in certain strongly-correlated fractional topological systems, such as fractional quantum Hall systems~\cite{Stormer1999} or $Z_2$ spin liquids~\cite{Wen2017}, a low-energy excitation
%are ``smaller" than the fundamental building blocks, i.e., the quantum numbers that they carry 
is just a fraction of the fundamental building blocks, and this phenomenon is known as fractionalization. More specifically, the definition of fractional excitations involves five criteria. (1) ``Integer" excitations need to be defined, i.e., the system needs to obey certain quantization condition, such that excitations can be classified by certain integer quantum number (e.g., charge). 
(2) An integer excitation then  ``breaks up" into multiple pieces. Most importantly, the interactions between these pieces need\KS{s to} be weak and decay to zero when they are separated faraway in distance, known as {\it deconfinement}. In quantum systems, deconfinement is a highly nontrivial requirement, because it is usually impossible to break a quantum particle, e.g., an electron. In classical physics, it is often possible to partition an object. However, such partition in classical physics usually cannot meet the next criterion. (3) Equal partition has to be enforced as we split the integer excitation. For example, if a charge $1$ integer excitation splits into $2$ equal parts, each part is a fractional excitation with charge $1/2$. Such equal partition is natural in quantum systems, but a nontrivial requirement in classical systems. Furthermore, two more criteria need to be enforced to ensure that these fractional excitations cannot be trivially mapped back to integer ones: (4) a fractional excitation must be a topological object, which cannot be created by any local deformations, and (5) these fractional excitations must exhibit novel properties impossible for any integer ones, such as braiding~\cite{Nayak2008}.

In this letter, we show that minimal-surface elastic plates support fractional low-energy excitations. Due to the presence of the minimal surface associate family, such systems exhibit two types of soliton configurations: integer and half-integer, in strong analogy to the quantum integer and fractional solitons in the one-dimensional (1D) dimerized spin chains of Haldane~\cite{Haldane1982}. In addition, we demonstrate that the classical system and the quantum spin chain share the same low-energy effect theory (compact sine-Gordon), and in both systems, fractionalization is induced by a $Z_2$ symmetry. As a result, this classical version of fractional excitations share identical physical properties as their quantum counterpart. For example, integer solitons are conventional and could be created via local deformations, but once it splits into two fractional solitons, each fractional soliton is a topological excitation, robust against any local perturbations.  

These fractional excitations exhibit exotic mechanical properties, including braiding which is general to fractional excitations, and holographic property and diode-like torque-rotation response which are unique to these minimal surface plates.  These novel properties may find broad applications as mechanical metamaterials for chirality flipping, mode conversion, wave rectification, impact mitigation, and mechanical logic circuits.

%%%%%%%%%%%%%%%%

\noindent{\it 2D non-Euclidean plates}--- The elastic energy of a 2D sheet with a non-Euclidean reference metric $g_0$ is composed of two parts $E=E_s+E_b$ for stretching ($E_s$) and bending ($E_b$) energies~\cite{EFRATI2009762,sharon2010mechanics}. 
The stretching energy depends on the first fundamental form (i.e., the metric tensor $g$) of the manifold 
\begin{align}
E_s=h \int d\mathcal{A} \left \{ \frac{B_0-G_0}{2} \tr(g-g_0)^2+ G_0 \tr[(g-g_0)^2] \right\} ,
%E_s=h \int dr{\sqrt{\det g_0}}\large\{ \frac{B_0-G_0}{2} & \tr(g-g_0)^2
%\nonumber\\
%&+ G_0 \tr[(g-g_0)^2]\large\}
\label{eq:Es_2D}
\end{align}
where $h$ is the thickness of the sheet and the elastic moduli are $B_0=\frac{9 B G}{4(3B+4G)}$ and $G_0=G/4$ with $B$ and $G$ being the 3D bulk and sheer moduli of the material respectively. $E_s$ is minimized if $g=g_0$.
The bending energy depends on the second fundamental form (i.e., the curvature tensor) $b$. In this study, we focus on 2D non-Euclidean \emph{plates}, i.e.  thin sheets homogeneous along the thickness direction.  For a 2D plate, the bending energy takes the following form
\begin{align}
E_b=h^3  \int d\mathcal{A}\frac{G}{12}\left[ \frac{8(3B+G)}{3B+4G} H^2 -2 K \right],
%E_b=h^3  \int dr{\sqrt{\det g_0}}\frac{G}{12}\left[ \frac{8(3B+G)}{3B+4G} H^2 -2 K \right],
\label{eq:Eb}
\end{align}
where $H$ and $K$ are the mean and Gaussian curvature respectively (i.e., the trace and determinant of the curvature tensor $b_i^j$).
Because $E_s \propto h$ and $E_b\propto h^3$, $E_s$ is the dominant part in the small thickness limit $h\to 0$. 

Here, we highlight one important symmetry property of 2D plates: the elastic energy  is invariant if the curvature tensor flips sign ($b \to -b$), which is a $Z_2$ symmetry. This $Z_2$ symmetry originates from the fact that the two sides of a plate are fully equivalent, and thus the transformation $b \to -b$, which is equivalent to flip the two sides of the sheet is a symmetry operation that preserves the elastic energy. In Eq.~\eqref{eq:Eb} this symmetry is reflected by the fact that $E$ is an even function of $H=b^i_i$, and this $Z_2$ symmetry plays a crucial role for fractional excitations.

%%%%%%%%%%%%%%%%

\begin{figure}[t]
	\centering
\includegraphics[width=1\columnwidth]{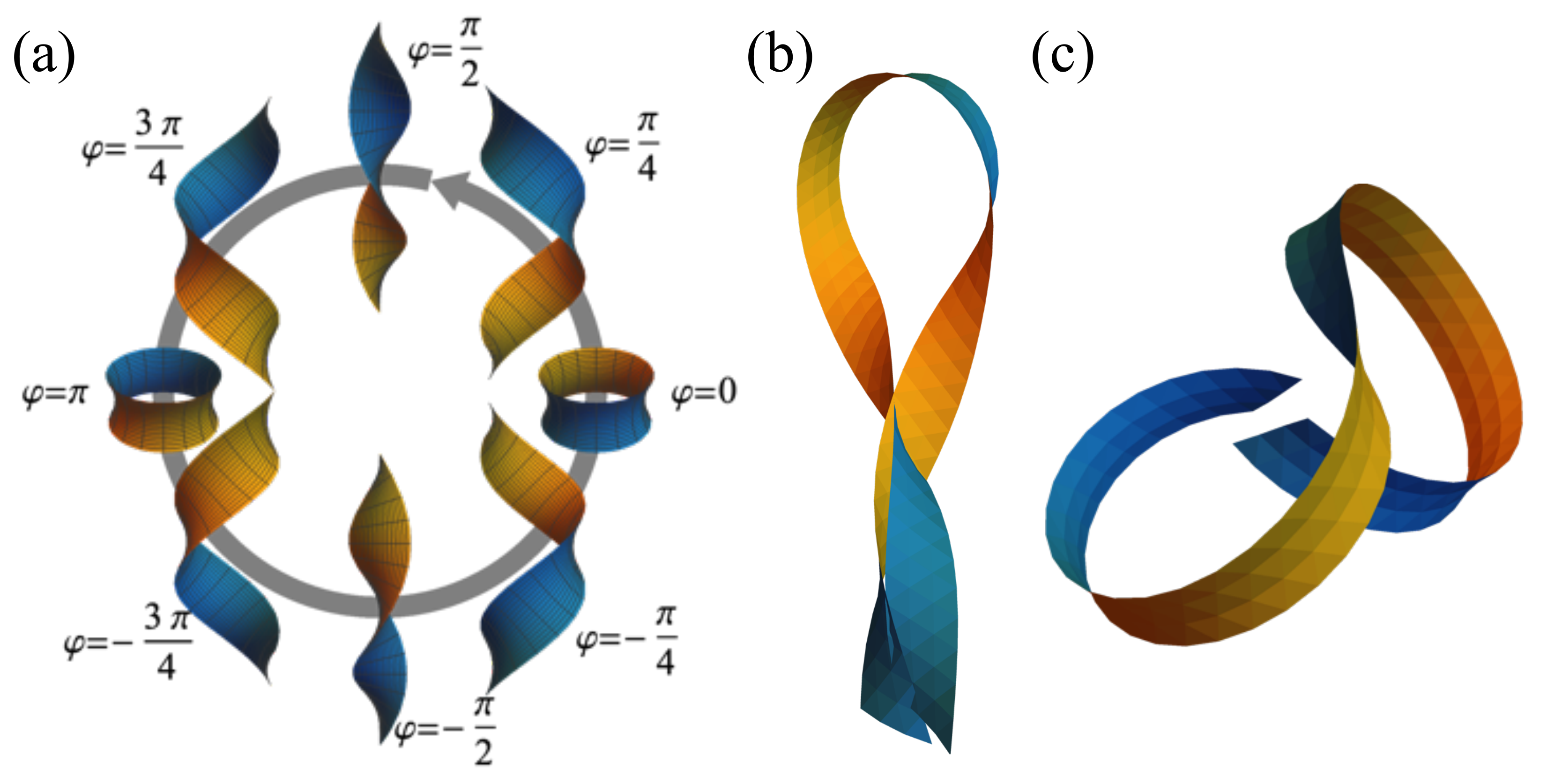}
	\caption{(a) The helicoid-catenoid associate family. (b-c) soliton configurations from finite-element analysis for ribbons where a helicoid (b) and a catenoid (c) are the ground states respectively.}
	\label{fig:associate:family}
\end{figure}

%%%%%%%%%%%%%%%
\noindent {\it Minimal surfaces and low-energy effective theory}---Minimal surfaces are 2D surfaces that minimizes its area locally, characterized by a vanishing mean curvature $H=0$.
In this letter, we focus on 2D plates whose target metric tensor ($g_0$) is that of a minimal surface. In this case, minimization of elastic energy $E$ gives $g=g_0$ and $H=0$~\cite{Efrati2011}.  
%By minimizing the elastic energy above, it is easy to realize that the ground state configuration here is a minimal surface with $g=g_0$ and $H=0$. 
However, this doesn't uniquely determine one ground state configuration. Instead, there exist infinitely many minimal surfaces with $g=g_0$ and $H=0$ and all these configurations are degenerate ground states of $E$ (where the only nonzero term is $-2K$ which is fully determined by $g_0$ and thus is a constant)~\cite{Levin2016}. 
This set of minimal surfaces, which share the same metric tensor, are called an ``associate family"~\cite{dillen2000handbook}. It has been known that  minimal surfaces in an associate family can be labeled by a phase angle $\varphi$. As we vary $\varphi$, minimal surfaces in this associate family deform smoothly into each other. As $\phi$ is increased by $2\pi$, the surface returns to its original configuration. One such example, helicoid-catenoid associate family, is shown in Fig.~\ref{fig:associate:family}(a)~\cite{supplement}.  

In summary, the associate family that a minimal surface plate belongs to defines a \emph{soft mode} of this plate, where we can deform the plate with zero elastic-energy cost to the leading order (up to $O(h^3)$). This soft mode dominates low-energy deformations of such plates. 

In particular,  we consider a long ribbon of a 2D minimal surface plate. 
%(the length much larger than the width). 
Here, low-energy excitations can be characterized by a slowly varying $\varphi$ along the ribbon direction $\varphi(v)$, where $v$ is the coordinate along the ribbon. In an ideal minimal surface plate, because all configurations in the associate family have the same energy, the elastic energy  take the following form to the leading order $E=\int d v \left[ (\partial_v \varphi)^2\right]$, i.e., energy cost from inhomogeneity.
However, in reality, due to the finite thickness and other deviations from the ideal 2D limit, different configurations in the associate family may have some small energy difference, and thus an additional term  arises $E=\int d v \left[ (\partial_v \varphi)^2+V(\varphi)\right]$~\cite{supplement}. For a generic 2D minimum surface, $V$ must be a periodic function with $V(\varphi)=V(\varphi+2 \pi)$ due to the periodic structure of the associate family. For simplicity, here we will take the lowest Fourier harmonic $V(\varphi)=\gamma \cos (\varphi-\varphi_{0})$, but it must be emphasized that the same qualitative features we discuss below survive even if more complicated $V(\varphi)$ is considered. 
As a result, the elastic energy now takes the form of a sine-Gordon theory, which supports soliton solutions. Here, we define the soliton charge as $\Delta\varphi/(2\pi)$, where $\Delta\varphi$ measures the change of $\varphi$ across a soliton. Due to the periodicity $V(\varphi)=V(\varphi+2 \pi)$,  it is easy to verify that the soliton solution for this sine-Gordon elastic energy has $\Delta\varphi=2\pi$ and thus the soliton charge is $1$. Therefore, they will be called  integer excitations (i.e., integer solitons). This quantization is due to the periodic structure of the associate family.

\noindent {\it Fractional excitations}---
In 2D plates, the $Z_2$ symmetry discussed early on enforces a nontrivial constraint. In a minimal surface associate family, this $Z_2$ transformation ($b\to -b$) corresponds to $\varphi \to \varphi+\pi$. Thus, it implies that the elastic energy  remains invariant under $\varphi \to \varphi +\pi$. As a result, we must also have $V(\varphi)=V(\varphi+\pi)$, i.e., the periodicity of the function  $V(\varphi)$ is reduced  from $2\pi$  to $\pi$. As a result, if we still focus on the lowest harmonics in $V(\varphi)$, the elastic energy becomes
\begin{align}
E=\int d v \left\{(\partial_v \varphi)^2+  \gamma \cos  \left[2 (\varphi-\varphi_{0})\right]\right\}
\label{eq:sine_gordon}
\end{align}
where an extra factor of $2$ arises in the $\cos$ function. With this extra factor of $2$, the soliton configuration of this sine-Gordon theory has $\Delta\varphi=\pi$, and thus the soliton charge is $1/2$. These are the fractional solitons.

This mechanism of symmetry-induced fractionalization is identical to the fractional solitons in Haldane's dimerized spin chain, where fractional spin-1/2 solitons arise from a $Z_2$ symmetry (i.e., translation by an odd integer times the lattice constants)~\cite{Haldane1982}. This physics is also in strong analogy to nematic liquid crystals, where the molecules (and the order parameter) are invariant under a $\pi$ rotation, and this $Z_2$ symmetry then results in fractional topological defects in nematic liquid crystals, i.e., disinclinations or disinclination lines, which can be viewed as half of a vortex or a vortex line~\cite{degennes1993, Lubensky2000}.

%%%%%%%%%%%%%%%%%%
Guided by the insight obtained from the low-energy effective theory, we perform finite element analysis for helicoid- and catenoid- ribbons as an example to verify the existence of fractional solitons as their low energy excitations.
%and their studied low-energy excitations, where fractional/integer solitons are indeed observed.
In particular, we simulate a narrow ribbon with $E=E_s + E_b$ as given in Eq.~(\ref{eq:Es_2D},\ref{eq:Eb}) with $g_0$ of the helicoid-catenoid associate family.  A small perturbation is added to $E_b$ to lift the infinite degeneracy of the ground states, favoring either the helicoid ($\varphi=\pm \pi/2$)  as ground states or the catenoid ($\varphi=0,\pi$) as ground states, corresponding to $\varphi_0=0$ and $\varphi_0=\pi/2$ in Eq.~\eqref{eq:sine_gordon} respectively~\cite{supplement}.
This simulation didn't enforce the excluded-volume condition, and thus the ribbon may intersect with itself.  Enforcing excluded volume or not doesn't change any qualitative conclusions.

From this finite-element analysis, we found that a fractional soliton is indeed a local energy minimum, as shown in Fig.~\ref{fig:associate:family}(b-c). For the case where the helicoid is the true ground state, the fractional soliton is the domain boundary between a left-handed (L) section of helicoid and a right-handed (R) one. For the case where the catenoid is the true ground state, the fractional soliton is also a domain boundary, across which the two sides (inside and outside) of the catenoid flips. The fact that $1/2$ soliton corresponds to a domain boundary is universally true for any 1/2 soliton in any minimal surface plates, as well as in dimerized quantum spin chains~\cite{Haldane1982}. Because it is a domain boundary, such fractional solitons cannot be created by any local deformations, in contrast to integer solitons, which are just regular local excitations in these systems and can be created or removed locally.

In particular, for the helicoid ground states, 
by minimizing the elastic energy, we find that such a domain structure
always bends the ribbon by nearly $180^\circ$, i.e., if we move along  the helicoid ribbon, each soliton excitation implies a sharp U-turn. The origin of this sharp turn is that $\varphi$ changes between $\pm \pi/2$ across the soliton, thus the soliton profile is characterized by a narrow section of a catenoid, which naturally turns the ribbon.
%The reason for this sharp turn can be understood by looking at a section of catenoid and trying to connect it with two helicoids with opposite chirality to its two ends. As shown in Fig.~\ref{fig:associate:family}, if one tries to smoothly connect them (by slowly varying $\varphi$ along the chain), a $180^\circ$ degree turn shall naturally arise. 
%This sharp U-turn associated with such a fractional soliton is the key for the formation of kinks in tangled phone-cords.

\begin{figure}[t]
	\centering
	\includegraphics[width=1\columnwidth]{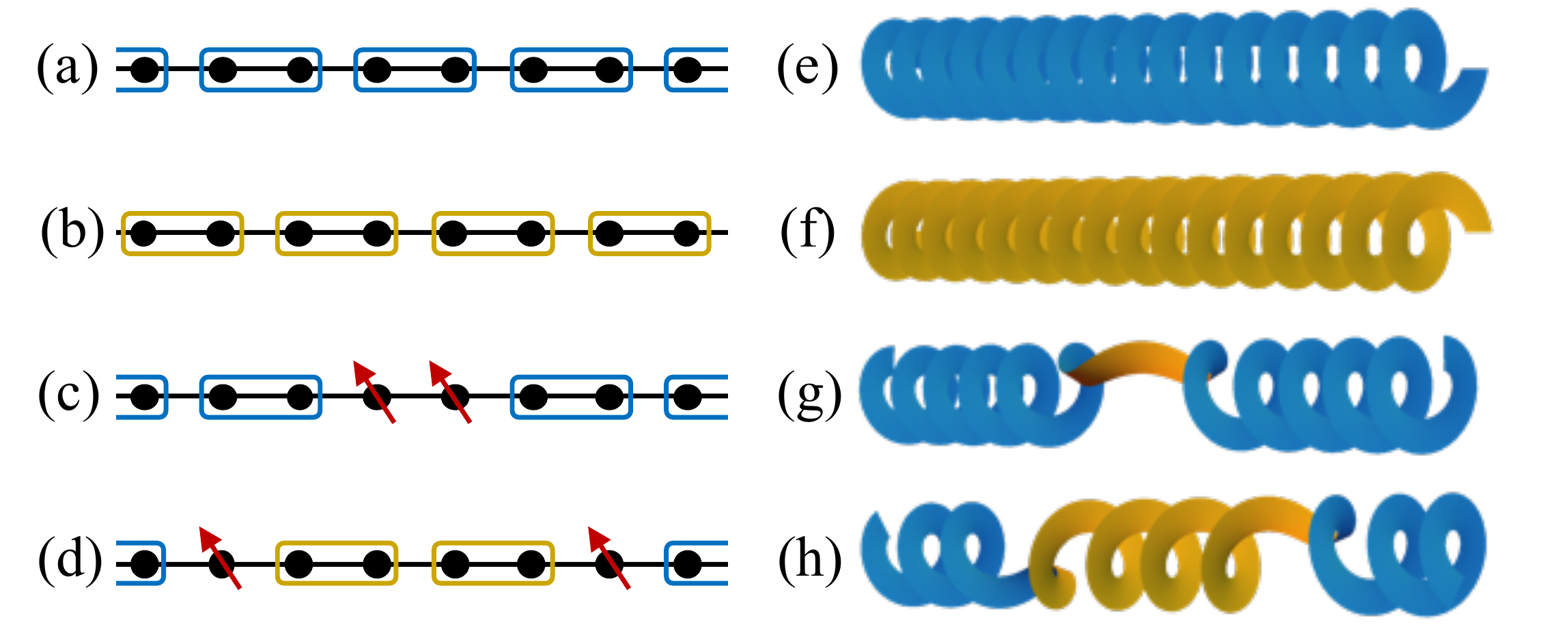}
	\caption{Analogous fractional solitons in  (a-d) a dimerized spin chain and (e-h) a helicoid ribbon. The quantum spin chain has two degenerate ground states (a) and (b). A spin-1 excitation can be created via local perturbations (c), which splits into two deconfined spin-1/2 excitations in (d). (e) and (f) show two degenerate ground states of a helicoid ribbon with opposite chirality. 
	(g) shows a local charge-1 soliton, which splits into a pair of fractional solitons in (d).}
	\label{fig:spin:chain}
\end{figure}

\noindent{\it Quantum-classical analogy and braiding}---
To set the stage for comparing these classical fractional excitations with their quantum counterparts, here we first provide a brief review of 1D dimerized spin chains and 2D $Z_2$ spin liquids. 
A 2D $Z_2$ spin liquid is one of the most important and well-studied fractional topological states (see e.g.~Refs.~\cite{Fradkin2013,Wen2017} and references therein).
The study of $Z_2$ spin liquids originates from Anderson's resonating-valence-bond (RVB) scenario~\cite{Anderson1973, Fazekas1974} in frustrated quantum spin systems and quantum dimer models~\cite{Kivelson1987, Rokhsar1988, Moessner2001}. 
This exotic quantum phase of matter is characterized by a topological Ising gauge theory and gives rise to deconfined fractional excitations, e.g., spinons which carry spin-$1/2$ 
but no charge~\cite{Read1991,Wen1991,Mudry1994, Senthil2000, Moessner2001b}. Later, an exactly sovable model with the same topological order was introduced, known as the toric code model of Kitaev~\cite{Kitaev2003}. A 1D dimerized spin chain (e.g., the Majumdar-Ghosh model~\cite{Majumdar1969}) does not show a $Z_2$ topological order, but it shares certain similar feature as the $Z_2$ spin liquids.

Here we start from the simpler 1D case by considering the 1D Majumdar-Ghosh model (spin-$1/2$ Heisenberg spins with frustrated nearest and next-nearest-neighbor anti-ferromagnetic couplings)~\cite{Majumdar1969}. This model has two dimerized ground states as shown in Fig.~\ref{fig:spin:chain}(a) and (b), where each box represent a spin singlet pair, known as a ``dimer".  One obvious excitation in such a ground state is to break a dimer, transferring a singlet into a triplet, which carries integer spin $S=1$. However, such a local excitation can fractionalize into two deconfined spin-1/2 fractional excitations, as shown in Fig~\ref{fig:spin:chain} (c) and (d), which are the fractional solitons as pointed out by Haldane~\cite{Haldane1982}.

One important and unique property of these fractional particles is that by moving such fractional particles around non-contractible loops, the global state of the entire system can be transformed  in a nontrivial way. One such example is ``braiding" (i.e., moving particles around each other), which play a crucial role in the understanding of fractional quantum Hall effects, Majorana modes and topological quantum computing~\cite{Nayak2008}. Here, instead of moving a fractional particle around another fractional particle, we take a different non-contractible loop, which reflects the same fundamental  principles.

As shown in Fig.~\ref{fig:spin:chain}(d), if we move the two aforementioned fractional excitations in the 1D Majumdar-Ghosh model away from each other, we flipped the ground state from (a) to (b). This phenomenon also arises in minimal-surface plates, as shown in Fig.~\ref{fig:spin:chain}(e-h). An integer soliton of charge-1 can be locally generated.  This soliton can split and turn into two charge-1/2 fractional solitons.  This pair of fractional solitons are deconfined, as the ribbon  between them is in ground state.  Moving this pair of fractional solitons away from each other flips the ribbon between R- and L- helicoids.

\begin{figure}[t]
\includegraphics[width=1\columnwidth]{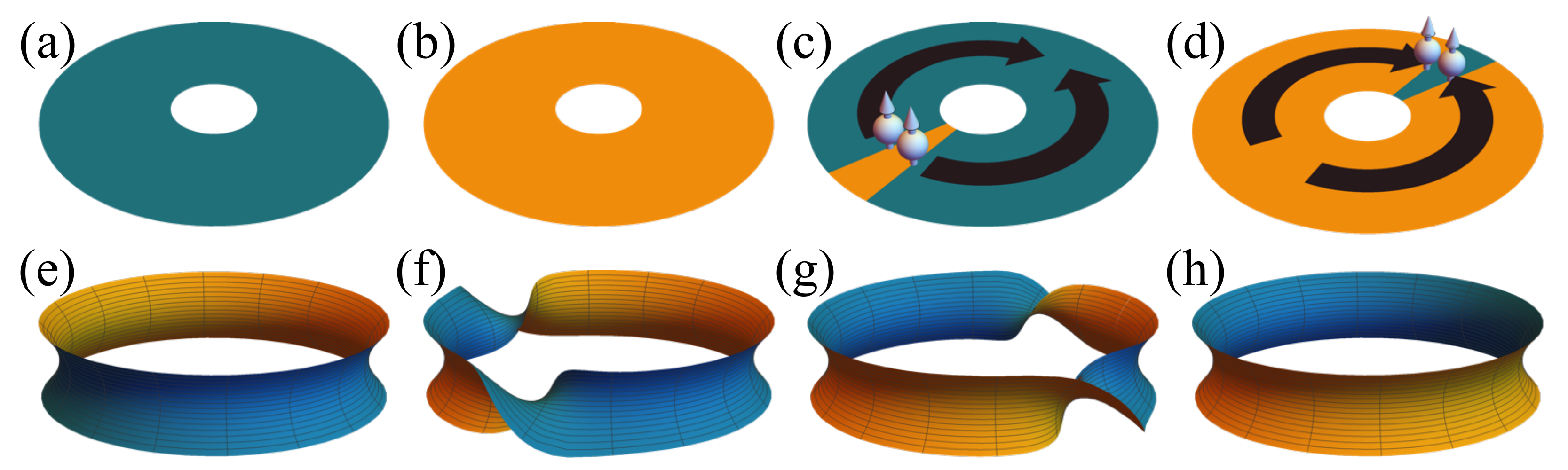}
	\caption{Fractional excitations and topological degeneracy. (a-d) A quantum $Z_2$ spin liquid defined on an annulus, which has two degenerate ground states (a) and (d) due to topological degeneracy. Via local perturbations, a spin-1 excitation is introduced to the first ground state, which can split into two spin-1/2 fractional excitations as show in (b). If these two fractional excitations are moved around the annuls and then annihilated with each other (c), the system turns into the other ground state (d). (e-h) A catenoid with the same geometry setup shows the same property. The two degenerate ground states correspond to swap the two sides of the 2D manifold (e and h). One can create two charge-1/2 solitons (f) and move them around the catenoid (g) before annihilate them, flipping the catenoid to the other ground state. }
	\label{fig:spin:liquid}
\end{figure}

For a 2D $Z_2$ spin liquid, a similar phenomenon arises~\cite{fradkin2013}, where moving a pair of fractional excitation around an annulus flips the topologically degenerate ground states, as shown in Fig.~\ref{fig:spin:liquid}. This is analogous to  the motion of fractional excitations in the a catenoid.

%%%%%%%%%%%%%%%%%%%%%%%%%%%%%%%%%%
%%%%%%%%%%%%%%%%%%%%%%%%%%%%%%%%

\noindent{\it Holographic property}---
In addition to the analogy with their quantum counterparts, fractional solitons in a  helical ribbon have certain unique features, not generally expected for fractional excitations.
One such example is that these solitons are \emph{holographic}, which means that if there is only one charge-1/2 soliton in a helicoid, we can pin-point and control its location via 
controlling the two ends of the helicoid.
This is because this soliton  is the domain boundary between the L and R  sections. For a helicoid of length $l$, we define the L section length to be $x$, and thus the R section length is $l-x$. For simplicity,  we ignore the width of the soliton. The helicity of the whole ribbon, defined as the net number of R twist, is then $(l-2x)/\lambda$ where $\lambda$ is the pitch of the helicoid.  This directly relates helicity to the position of the soliton.  Thus, by twisting the two ends of the ribbon relative to one another, one can change helicity and the position of the soliton holographically. 
%We can define the total helicity of the entire ribbon as $(l-2x)/\lambda$ where $\lambda$ is the pitch of the helicoid. This quantity describes how many times the ribbon twists, with R twists defined as positive. Thus the total helicity 
%This quantity directly connects helicity with the location of the fraction soliton $x$. 
%For an elastic ribbon, the helicity can be adjust by twisting the two ends of the ribbon in the opposite direction and each $2\pi$ twist increases/decreases the helicity by $1$, which moves the fractional soliton by one pitch $\lambda$. 
This holographic control is not a general property of fractional solitons, but a special feature for solitons in helicoids. 
In addition, this holographic property also provides a natural way to generate these fractional excitations. If we twist the two ends of a helicoid such that the helicity decreases from the ground state value ($l/\lambda$), this process will  create a fractional soliton (i.e. a non-zero $x$) to reduce energy. 
%As will be discussed in the next section, this is the mechanism how solitons in telephone cords is generated.

\begin{figure}[h]
	\centering
	\includegraphics[width=1\columnwidth]{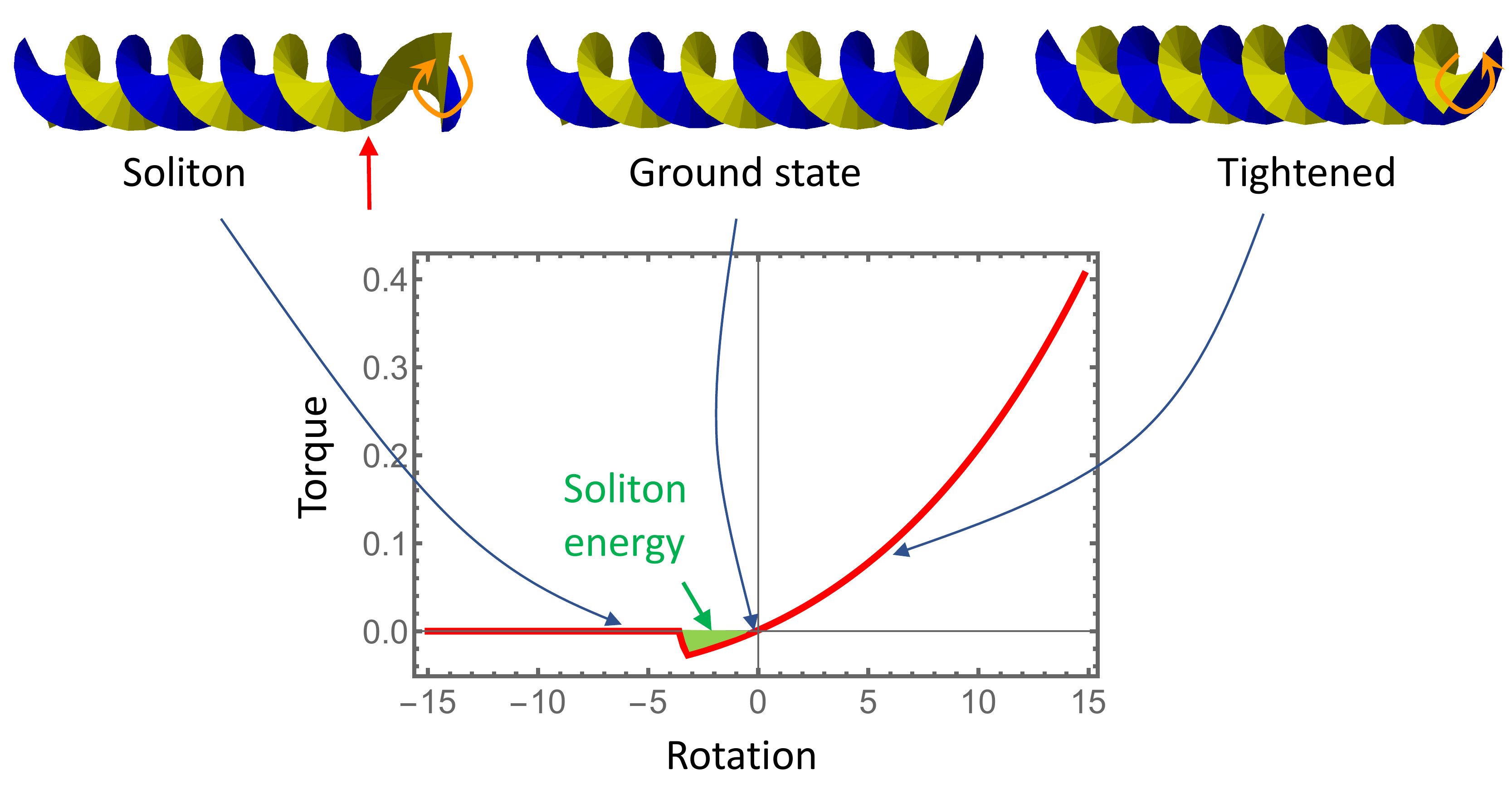}
	\caption{Diode-like torque-rotation response.  
	Three representative configurations are shown for the ground state (middle), a state with counterclockwise rotation (orange arrow) where the helicoid is tightened (right), and a state with clockwise rotation where a soliton (red arrow) is generated. The helicoid is R on the left and L on the right of the soliton. 
	%The green area shows the energy of generating a soliton.
	%(a) Illustration of clockwise and counter-clockwise rotation of one end of a helical ribbon.  (b) Simulated torque-rotation response of this ribbon, resembling the IV characteristics of a diode. 
	}
	\label{FIG:Diode}
\end{figure}

\noindent{\it Diode-like torque-rotation response}
The holographic property of this fractional soliton gives rise to unusual mechanical response.  One prominent example is that when one end of the ribbon is fixed and the center-line of the ribbon is confined to be straight (e.g., by embedding a stiff rod), the torque-rotation relation at the opposite end strongly resembles the current-voltage (IV) characteristics of a diode.  
We simulated this effect assuming an elastic energy of the form in Eq.~\eqref{eq:sine_gordon}, and the results are shown in Fig.~\ref{FIG:Diode}. When counterclockwise rotation is applied to the end of an R helicoidal ribbon, it tightens the ribbon and leads to a linear torque-rotation response.  In contrast, when clockwise rotation is applied to the end of this ribbon, it generates a soliton, which turns the R helicoid into an L helicoid.  At very small counter-clockwise rotation the response is still linear (which homogeneously loosens the helicoid), but as the rotation increases, a small barrier is overcome and the torque vanishes, as further rotation just moves the soliton to the left, where the elastic energy of the ribbon stays constant.  This barrier (green area in Fig.~\ref{FIG:Diode}) equals to the energy of one soliton as given by Eq.~\eqref{eq:sine_gordon}.  This strong asymmetry resembles the IV characteristics of a diode, where voltage of different directions generates currents of dramatically different amplitudes.  

Furthermore, this system exhibits convenient programmability by placing the soliton at different positions in the ribbon.  This correspond to shifting the torque-rotation curves horizontally, leading to programmed torque response. This effect can potentially apply to a broad range of problems such as wave rectification, impact mitigation, and mechanical logic circuits.

\noindent{\it Conclusion and discussion}---
We demonstrate that due to minimal surface associate families, non-Euclidean elastic plates can support low energy fractional soliton excitations that strongly resemble fractional quantum excitations.  These fractional solitons are highly robust and can not be locally created or destroyed.  They exhibit novel mechanical properties such as holographic control and diode-like torque-rotation response.

This type of non-Euclidean plates can be realized in experiments through various techniques of metric control, such as stimuli responsive gels, strain engineering, halftone and gray-scale 3D printing~\cite{sharon2010mechanics,gladman2016biomimetic,chen2012nonlinear,Kim1201}.  The unique holographic control and diode-like nonlinear elastic response may open the door to novel mechanical metamaterials.

In addition, the fractional soliton in the case of helicoid elastic ribbons share a lot of similarities with various types of kinks and perversions between domains of different handedness in other helical structures such as tendrils on climbing plants~\cite{Goriely1998}, intrinsically curved rods~\cite{domokos2005multiple}, self-assembled structures of Janus colloidal particles~\cite{Chen2011supra},
elastic bi-strips~\cite{liu2014structural}, helical strings~\cite{Nisoli2015}, and minimal surface liquid films~\cite{Machon2016}.  In this paper we reveal their unexpected link with fractional quantum excitations.
%show how this type of solitons arises  in narrow elastic ribbons and are fractional excitations flipping the ribbon between ground states that belong to the same minimal surface associate family.  We further demonstrate that they share the same topological description as fractional excitations in $Z_2$ spin liquids.

\noindent{\it Acknowledgement}---
This work is supported in part by National Science Foundation (NSF-EFRI-1741618) and the Office of Naval Research (MURI N00014-20-1-2479).

%\bibliography{isostaticity}

%merlin.mbs apsrev4-1.bst 2010-07-25 4.21a (PWD, AO, DPC) hacked
%Control: key (0)
%Control: author (8) initials jnrlst
%Control: editor formatted (1) identically to author
%Control: production of article title (-1) disabled
%Control: page (0) single
%Control: year (1) truncated
%Control: production of eprint (0) enabled
%

\vspace{2cm}

\centerline{\LARGE \bf{Supplementary Materials}}

\section{Minimal surfaces}
\label{app:sec:minimal:surface}
To make the manuscript self-contained, here we provide a brief review about basic concepts and properties of minimal surfaces. Minimal surfaces are 2D surfaces that locally minimize their area, which is equivalent to requiring these surfaces to have zero mean curvature.

\subsection{Weierstrass-Enneper parameterization}
Mathematically, minimal surfaces have a deep and fundamental connection with complex analysis. 
It is know that all minimal surfaces can be represented using the Weierstrass-Enneper parameterization
\begin{align}
R_1&=\Re  \int \mathfrak{f}  (1-\mathfrak{g}^2)/2 \;\; dz \\
R_2&= \Re \int i \mathfrak{f} (1+\mathfrak{g}^2)/2 \;\; dz  \\
R_3&=\Re \int \mathfrak{f \; g} \;\;dz
\end{align}
where $\Re$ represents the real part. $\mathbf{R}=(R_1, R_2, R_3)$ is the 3D coordinate of the target space, while the 2D coordinate of the original space (i.e. the 2D manifold) $\mathbf{r}=(x,y)$ is represented by the  complex variable $z=x+i y$. $\mathfrak{f}$ and $\mathfrak{g}$ are complex functions of $z$, where $f$ is holomorphic and $g$ is meromorphic. In complex analysis, holomorphic means that a function is analytic with well defined Taylor expansions for every point in a domain, while meromorphic is a slightly weaker condition, which is similar to an analytic function but can contain a set of isolated singular points (poles). It is easy to verify that the 2D manifold defined by $\mathbf{R}(\mathbf{r})$ has zero mean curvature and thus is a minimal surface.

In the Weierstrass-Enneper parameterization, an associate family is represented by a phase factor $e^{i\varphi}$. As can be easily verified, by multiplying a constant phase factor to the function $\mathfrak{f}$ , we obtain a family of minimal surfaces via the Weierstrass-Enneper parameterization  
\begin{align}
R_1&=\Re\; e^{i \varphi} \int \mathfrak{f}  (1-\mathfrak{g}^2)/2 \;\;dz \\
R_2&= \Re\; e^{i \varphi}\int i \mathfrak{f} (1+\mathfrak{g}^2)/2 \;\;dz  \\
R_3&=\Re \;e^{i \varphi} \int \mathfrak{f\; g} \;\;dz
\end{align}
All minimal surfaces in this associate family share the same metric tensor $g$ and the same mean curvature ($H=0$), and they can be evolved smoothly into each other via 
adiabatically varying the value of $\varphi$. Because our elastic energy $E=E_s+E_b$ only depends on $g$ and $H$, for a 2D plate with minimal-surface ground state, all configurations in the associate family are degenerate ground states and there exits a floppy mode that deforms these ground states into each other smoothly without 
any energy cost.

From equations shown above, it is easy to realize that under the transformation $\varphi \to \varphi+\pi$, $\mathbf{R}\to-\mathbf{R}$, which flips the chirality. 

\subsection{The helicoid-catenoid family}
For minimal surfaces in the helicoid-catenoid family, the Weierstrass-Enneper parameterization takes the following form
\begin{align}
\mathfrak{f}= i e^{-z} \;\;\; \mathrm{and}\;\;\; \mathfrak{g}=-i e^{z}
\end{align}
Thus
\begin{align}
R_1&=- \cosh x \sin y \cos \varphi - \sinh x \cos y \sin\varphi \\
R_2&= \cosh x \cos y \cos \varphi- \sinh x \sin y \sin\varphi   \\
R_3&=x \cos\varphi - y \sin \varphi 
\end{align}
Under a coordinate transformation $u=\sinh x$ and $v=y$, we get
\begin{align}
R_1&=-\sqrt{1+u^2} \sin v \cos \varphi - u \cos v \sin\varphi 
\label{eq:cooridnate_app1}
\\
R_2&=\sqrt{1+u^2} \cos v \cos \varphi - u \sin v  \sin\varphi 
\label{eq:cooridnate_app2}
\\
R_3&= \arcsinh u \cos\varphi - v \sin \varphi 
\label{eq:cooridnate_app3}
\end{align}
For $\varphi=\mp \pi/2$, we have helicoids with left/right handness
\begin{align}
\mathbf{R}=\pm(u \cos v, u \sin v,v)
\end{align}
For $\varphi=0$ or $\pi$, catenoids are obtained with
\begin{align}
\mathbf{R}=\pm (-\sqrt{1+u^2} \sin v, \sqrt{1+u^2} \cos v, \arcsinh u)
\end{align}
Other values of $\varphi$ gives other minimal surfaces in this associate family.

\section{Higher order terms}
\label{app:sec:higher:order}
As mentioned in the main text, for a 2D plate that satisfies the minimal-surface criterion, the system has a floppy mode and thus infinitely many degenerate ground states, 
i.e. all minimal surfaces in the corresponding associate family. In a real material, such a infinite ground-state degeneracy will in general be lifted by 
higher order terms in the elastic energy. 
For a ribbon, this will result in two degenerate ground states, connected with each other by the $\varphi \to \varphi+\pi$ transformation as shown
in the main text. In this section, we demonstrate one example of such higher order terms, which favor the helicoid- or catenoid- ground states.

As shown in main text, for an isotropic (or nearly isotropic) material, the bending energy depends on the the mean curvature $H$ and the Gaussian curvature $K$.
In a neighborhood of any non-singular point of a smooth 2D manifold, an orthogonal coordinate $(u,v)$ always exists, under which the metric tensor becomes diagonal
\begin{align}
g=
\begin{pmatrix}
E(u,v) & 0\\
0 & G(u,v)
\end{pmatrix}.
\end{align}
In such a orthogonal coordinate, we  can define an $\tilde{h}$ matrix,
\begin{align}
\tilde{h}=
\begin{pmatrix}
l(u,v) & m(u,v)\\
m(u,v) & n(u,v)
\end{pmatrix}
\label{app:h_tilde}
\end{align}
where $l=L/E$, $n=N/G$ and $m=M/\sqrt{E G}$, and
\begin{align}
h=
\begin{pmatrix}
L(u,v) & M(u,v)\\
M(u,v) & N(u,v)
\end{pmatrix}
\end{align} 
is the second fundamental form.

Using the $\tilde{h}$ matrix defined in Eq.~\eqref{app:h_tilde}, the bending energy can be written as
\begin{align}
E_b=&\int d\mathcal{A}
%\int dr{\sqrt{\det g_0}}
\left[ D_1 H^2 - D_2  K \right] \nonumber\\
=&\int d\mathcal{A}
%\int dr{\sqrt{\det g_0}}
\left[ \frac{D_1}{4} (l+n)^2 - D_2  (l n-m^2) \right]
\label{eq:Eb:app}
\end{align}
where $D_1$ and $D_2$ are coefficients shown in the bending elastic energy defined in the main text. 
Here, we utilized the fact that $H=\tr{\tilde{h}}/2$ and $K=\det \tilde{h}$.
This bending energy is isotropic, i.e., the energy cost is identical no matter we bend along the main axis ($u$ or $v$) or the diagonal direction ($u+v$ or $u-v$).

For a real 2D plate with $D_{4h}$ symmetry (e.g. materials with a tetragonal lattice) or lower symmetries, 
the bending term is no longer isotropic, and thus extra terms become allowed, such as
\begin{align}
\delta E_b=\int d\mathcal{A}
%\int dr{\sqrt{\det g_0}} 
\left[   \frac{\delta D}{4} (l^2+n^2)\right]
\label{eq:deltaEb:app}
\end{align}
where $\delta D$ is a coefficient, which describes the anisotropy between the main-axis and the diagonal directions. 
In this manuscript, we focus on nearly isotropic systems, and thus we will always assume that $\delta D \ll D_1$ and   $\delta D \ll D_2$.
For positive (negative) $\delta D$, this term implies that it is harder (easier) to bend the plate along the main-axis direction, in comparison to diagonal.
With $\delta D>0$, this term lifts the infinite degeneracy of the narrow ribbon and selects the L and the R helicoids as the real ground states.
%, i.e. the left- and right-handed helicoids.
If $\delta D<0$, the ground states are the two catenoids.
In the sine-Gordon description, this term (and other similar terms) gives rise to the cosine terms of higher order. 
\\

\section{Finite element simulation}
In the simulation, 10-node triangular elements are utilized. The shape function of such an element preserves the three-fold rotational symmetry, which help minimizing 
anisotropy induced by the shape function. The entire ribbon is composed of $60\times 4$ nodes. The elastic moduli are set to (in arbitrary units) 
$h B_0 = 6 \times 10^5$ and $G_0 =B_0/2$. The bending stiffness 
$D_1=\frac{2 h^3 G (3B+G)}{3(3B+4G)}$ is set to $ 2.4 \times 10^3$. 
We also added a small perturbation to favor the helicoid (or catenoid) ground states [as shown in Eq.~\eqref{eq:deltaEb:app}], 
whose coefficient is set to $\delta D=0.02 D_1$ for the helicoid ribbon and $\delta D=-0.01D_1$ for the catenoid.
For a minimal surface ribbon in the helicoid-catenoid associate family, the target matrix tensor $g_0$ can be set to
\begin{align}
g_0=
\begin{pmatrix}
1 & 0\\
0 & 1 + a_1 u +a_2 u^2
\end{pmatrix},
\end{align}
Here we use the coordinates shown in Eq.~\eqref{eq:cooridnate_app1}-\eqref{eq:cooridnate_app3}.
The control parameters are set to $a_1=0$ and $a_2= 2\pi/10$ in the numerical study.
All qualitative features that we observed are insensitive to microscopic details and remain stable as we vary the control parameters and the system size.

\section{Simulation of torque-rotation curve}
In this section we summarize the simulation that gives the  torque-rotation curve (Fig.~3) in the main text.

We start from the sine-Gordon low energy theory of the ribbon, and adapt this energy for the helicoidal ribbon.  More specifically we take the discretized energy of a helicoidal ribbon to be
\begin{align}
    E = \sum_{i=1}^{L} \left(\alpha_i^2 - \bar{\alpha}^2\right)^2 + \sum_{i=1}^{L-1} \kappa \left(\alpha_{i+1} - \alpha_{i}\right)^2 ,
\end{align}
where $i$ labels discrete sections along the long-axis of the ribbon, and $\alpha_i$ is the twisting angle at section $i$.  Here $\bar\alpha$ is the  twisting angle of the degenerate ground state helicoids.  The first term is a double-well potential, where the L and R helicoids are the two degenerate ground states $\pm \bar\alpha$.  The second term is the bending energy, where changing twisting angle from one section to the next costs energy.  $\kappa$ is the bending stiffness.

This equation can be viewed as a simplified version of Eq.~(3) in the main text around the helicoid ground state.  
This angle $\alpha$ can be related to the phase angle $\varphi$ of the minimal surface associate family.

To obtain the torque-rotation curve, we minimize $E$ at given 
\begin{align}
    \Phi=\sum_{i=1}^{L} \sin (\alpha_{i+1}),
\end{align}
which is the rotation of the right end ($i=L$) of the ribbon when the left end ($i=0$) is held fixed.  Note that here we sum $\sin (\alpha_{i+1})$ instead of $\alpha_i$ as we consider isometric deformations of the ribbon, which fixes the arclength of the ribbon's edge instead of the height of each section along the long-axis.

The minimization gives the shape and energy $E$ of the ribbon.  From the increment of the energy as $\Phi$ changes  we calculate the torque.

The coefficients we take for the simulation are given by $\bar\alpha=0.1\pi$, $\kappa=2\bar\alpha^2$, $L=80$.  As a result, the whole ribbon at the ground state is $4$ times the pitch length.

\end{document}